\documentclass[aps,twocolumn,showpacs]{revtex4}
\usepackage{graphicx}
\usepackage{amssymb}
\usepackage{bm}
\begin{document}


\newcommand{\re}[1]{(\ref{#1})}
\newcommand{\lab}[1]{\label{#1}}
\newcommand{\ci}[1]{\cite{#1}}
\renewcommand{\baselinestretch}{1.25}
\newcommand{\bfr}{\begin{flushright}}
\newcommand{\bfl}{\begin{flushleft}}
\newcommand{\efl}{\end{flushleft}}
\newcommand{\efr}{\end{flushright}}
\newcommand{\bc}{\begin{center}}
\newcommand{\ec}{\end{center}}
\newcommand{\be}{\begin{equation}}
\newcommand{\ee}{\end{equation}}
\newcommand{\bea}{\begin{eqnarray}}
\newcommand{\eea}{\end{eqnarray}}
\newcommand{\ba}{\begin{array}}
\newcommand{\ea}{\end{array}}
\newcommand{\edc}{\end{document}}
\newcommand{\ul}{\underline}
\newcommand{\ri}{\rightarrow\infty}
\newcommand{\li}{\leftarrow\infty}
\newcommand{\ra}{\rightarrow}
\newcommand{\la}{\leftarrow}
\newcommand{\ds}{\displaystyle}
\newcommand{\dsf}{\displaystyle\frac}
\newcommand{\dt}{\Delta{t}}
\newcommand{\il}{\int\limits}
\newcommand{\pal}{\partial}
\newcommand{\xxx}{{\it{X}}}
\newcommand{\bone}{{\bf 1}}
\newcommand{\gComment}[1]{}
\renewcommand{\gComment}[1]{\textcolor{red}{Gerardo: #1}}

\title{Anyon related correlations in two-dimensional Coulomb gases}


\author{B. Abdullaev$^1$, U. R\"{o}ssler$^2$, and C.-H. Park$^3$}

\address{$^1$
 Theoretical Physics Dept., Institute of Applied Physics,
Uzbekistan National University,
 Tashkent 700174, Uzbekistan}
\address{$^2$
Institute for Theoretical Physics, University of Regensburg, D-93040
Regensburg, Germany}
\address{$^3$
Research Center for Dielectric and Advanced Matter Physics,
Pusan National University, 30
Jangjeon-dong, Geumjeong-gu, Busan 609-735, Republic of Korea.}

\date{Received \today }


\begin{abstract}
In our recent paper (Phys. Rev. B {\bf 76}, 075403 (2007)), we have applied the anyon concept to derive an
approximate analytic formula for the ground state energy, which applies to two-dimensional (2$D$) Coulomb
systems from the bosonic to the fermionic limit. We make use of these results here to draw attention to
correlation effects for two special cases: the spin-polarized 2D fermion system and the charged anyon system
close to the bosonic limit. By comparison with quantum Monte-Carlo data (for the former) and exact results
obtained in the hypernetted-chain and Bogolyubov approximations (for the latter) we can conclude on correlation
effects, which have their origin in the bosonic systems and come into play by using the anyon concept. To
our knowledge, these correlations are not yet considered in the literature.
\end{abstract}


\pacs{71.10.Pm,\, 71.10.Ca,\, 71.10.Hf}

\maketitle

\newpage

The homogeneous electron gas, although it has become the testing ground of quantum mechanical many-body techniques
since long time and results of such calculations can be found in textbooks (see for example \ci{Kittel63} and
\ci{march}), does still contain yet unexpected puzzles. They are related to the correlation effects caused by the
particle-particle interaction, which can be treated so far (or principally) only in approximate ways. Here we
would like to draw attention to the two-dimensional ($2D$) systems, which have their prominent realizations in
semiconductor heterostructures and layered metallic systems such as the cuprate superconductors. These systems
attract much interest from experiment and theory due to the physics related with correlation effects
\ci{Ginliani,seidl,kivelson}.

With respect to the ground state energy in fermion systems, the Hartree-Fock (HF) approximation - originally
developed for high particle densities - provides the dominating contributions even for intermediate values of the
density parameter $r_s$. The difference between the exact ground state energy and its HF value is known as the
correlation energy. For $2D$ spin-polarized fermions, it has been demonstrated by quantum Monte-Carlo (MC)
calculations \ci{tanatar,attac}, that the correlation energy may be considered essentially as a small
correction to the HF energy for all values of $r_s$.

Calculations of the ground state energy of $2D$ Coulomb boson systems \ci{DePalo} have also been performed by
employing MC methods. As in \ci{tanatar,attac} these calculations were performed only for discrete
values of $r_s$ and an interpolation was made to yield an expansion in terms of powers of $r_s$. At small $r_s$ (the
high density limit), this interpolation formula reproduces the ground state energy of the $2D$ Coulomb boson gas (2DCBG)
obtained with the hypernetted-chain approximation \ci{Apaja}. The latter result follows also by using the Bogolyubov
approximation \ci{arm}.

In Ref.~\ci{arm} we have employed the anyon concept in order to find an approximate analytic formula for the
ground state energy of $2D$ Coulomb gases, which is capable of accounting for the aspects of fractional statistics
and brings together bosonic and fermionic aspects of the system. This formula was derived by using the
corresponding result for the harmonically confined 2$D$ Coulomb anyon gas \ci{aormn} and applying a regularization
procedure for vanishing confinement. Fractional statistics and Coulomb interaction have been taken into account
by introducing a function $f(\nu,r_s)$, which depends on both the statistics
and density parameters ($\nu$ and $r_s$, respectively), and was determined by fitting to the ground state energy
of the classical $2D$ electron crystal at very large $r_s$ (the 2$D$ Wigner crystal) and for very small $r_s$ (the
high density limit) to that of the 2DCBG
and to the HF energy of spin-polarized 2$D$ electrons. For the latter and at intermediate values of $r_s$
a comparison with HF (Ref.~\ci{rajagopal}) and MC (Refs.~\ci{tanatar,attac}) results has revealed significant
deviations, which are shown here in the enlarged scale of Fig.~1. It shows the correlation energy as a function of $r_s$
and exhibits a pronounced minimum with ${\cal{E}}_{c,min}(r_s\simeq 0.2)\sim -8~Ry$, where $Ry$ is Rydberg energy unit.
It appears essentially due to the minimum ${\cal{E}}_{0,min}$ of the ground state energy not only for fermions ($\nu = 1$) 
but for all $\nu\neq 0$, which is independent of the choice of $f(\nu,r_s)$ \ci{arm}.
The fitting of $f(\nu,r_s)$ to known ground state energies for high and low
density limits merely determines the $r_s$ value, at which the minimum  ${\cal{E}}_{0,min}$ appears.
We found also that as a function of the statistics parameter $\nu$ and close to the bosonic limit
($\nu\rightarrow 0$) the ground state energy exhibits for high densities ($r_s\rightarrow 0$) a minimum at finite
$\nu$, at which the energy diverges with $\sim -1/r_s$ faster than the energy of the 2DCBG for
$\nu = 0$ ($\sim-1/r_s^{2/3}$, Refs.~\ci{Apaja} and \ci{arm}).

Our expression for the ground state energy per particle derived in \ci{arm} (see Eq.~(33) of \ci{arm}) is entirely
determined by the function (see Eq.~(40) of \ci{arm})
\begin{eqnarray}
f(\nu,r_s)\approx \nu^{1/2}c_0(r_s)e^{-5r_s}&+&
\dsf{c_{BG}^{3/2}r_s/c_{WC}}
{1+c_1(r_s)c_{BG}^{3/2}r_s^{1/2}/c_{WC}} \nonumber\\
&+& \dsf{0.2 c_1(r_s)r_s^2
ln(r_s)}{1+ r_s^2} \,
\lab{ancorr1}
\end{eqnarray}
with
$c_0(r_s)=1+6.9943r_s+22.4717r_s^2$ and $c_1(r_s)=1-e^{-r_s}$. It fits to the expression of the
ground state energy of the classical $2D$ electron crystal \ci{bm}, $E_{WC}=-2.2122/r_s$ (for
$r_s\rightarrow\infty$). For $r_s\rightarrow 0$ it reproduces the ground state energy (expressed in $Ry$ energy units)
of the 2DCBG \ci{Apaja,arm}
\be
{\cal E}_0(\nu=0, r_s\rightarrow 0)=-
c_{BG}r_s^{-2/3} \, ,
\lab{ancorr2}
\ee
where
$c_{BG}=\frac{2\Gamma(-\frac{4}{3})\Gamma(\frac{5}{6})}{3\sqrt{\pi}}=1.29355$ and
the HF energy \ci{rajagopal} $E_{HF}=2/r_s^2- 16/3(\pi r_s)$
for spin-polarized 2$D$ electrons. In Eq.~\re{ancorr1} we used a constant $c_{WC}^{2/3}=2.2122$.

As seen from Fig.~1 and Fig.~2 of Ref.~\ci{arm} for the interval $0.7\le r_s< \infty $, the ground state energy
per particle (in $Ry$) can be well described by the formula
\be
{\cal E}_0(\nu, r_s \rightarrow \infty) =\dsf{c_{WC}^{2/3}
f^{2/3}(\nu,r_s)} {r_s^{4/3}}\left(-1+ \dsf{7\nu
f^{2/3}(\nu,r_s)}{3c_{WC}^{4/3}r_s^{4/3}}\right) \, ,
\lab{ancorr3}
\ee
which for low densities ($r_s \rightarrow \infty$)
or for $\nu < r_s$ is an approximate expression of the exact formula Eq. (33) of \ci{arm}. In either case, using
Eq.~(33) of \ci{arm} or Eq.~\re{ancorr3} (which is Eq.~(39) of \ci{arm}), the results fall below the MC data and the
corresponding interpolation formula by \ci{tanatar}. We allocate this deviation by observing that for small $r_s$
(but $r_s\ge 0.7$) the second term of $f(\nu,r_s)$ (Eq.~\re{ancorr1}), which is independent of $\nu$, becomes dominant and we
may replace $f(\nu,r_s)$ by its bosonic limit $f(\nu=0,r_s)$. Thus the dependence of the ground state energy
${\cal E}_0(\nu, r_s)$ on the anyon parameter $\nu$ is provided only by the factor $\nu$ in the second term of
Eq.~\re{ancorr3}. Therefore, the first term of Eq.~\re{ancorr3} with the common factor expressed by $f(0,r_s)$, which
is the expression for the boson ground state energy (Eq.~(36) of Ref.~\ci{arm}), dominates in ${\cal E}_0(\nu, r_s)$,
thus providing the
main bosonic contribution to the fermion ground state energy for the interval $0.7\le r_s\le 6$. For the interval
$0\le r_s\le 0.7$ the energy is expressed by exact formula Eq.~(33) of \ci{arm}.

In Fig.~1 we show the correlation energy for spin-polarized $2D$ fermions ($\nu=1$) obtained from the
exact formula Eq.~(33) of \ci{arm} for the ground state energy by subtracting the HF energy $E_{HF}$. The origin of
the deep minimum is connected with the second term of Eq.~\re{ancorr1}, discussed before, and can thus be ascribed to (the) bosonic
correlations, which become effective in our anyon approach and are absent in the fermionic descriptions.

\begin{figure}
\begin{center}
\includegraphics[angle=270,width=9.5cm,scale=1.0]{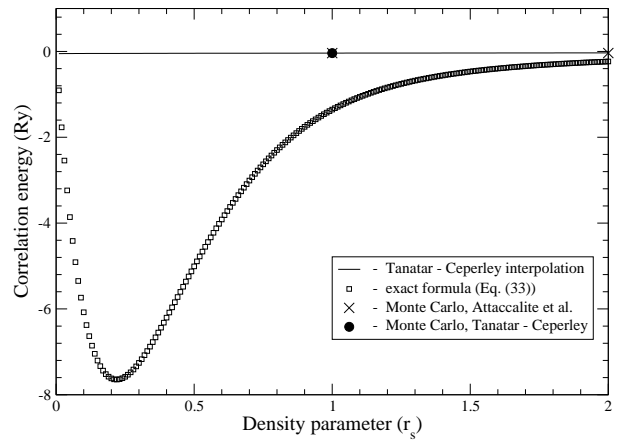}
\end{center}
\caption{Correlation energy as difference of the ground state energy (calculated from
exact formula Eq. (33) of Ref. ~\protect\ci{arm}) and the HF energy (open squares). MC data from
Refs. ~\protect\ci{attac} and  ~\protect\ci{tanatar} (cross and solid dot, respectively)
and interpolation between these MC data from Ref. ~\protect\ci{tanatar} (solid line) are given for comparison.
} \lab{fig1}
\end{figure}

The minimum of the ground state energy as function of $\nu$ found near the boson end ($\nu=0$) of the Coulomb
anyon gas occurs at $\nu_0=b_2^{3/4}c_{WC}r_s/b_1^{1/2}$ and takes the value
${\cal E}_{0 \, ,min}=-(4/5)b_2^{1/4}c_{WC}b_1^{1/2}/r_s$, where $b_1=1+2.441472r_s$ and $b_2=3/35$. It is
derived by using Eq.~\re{ancorr3} (because $\nu_0 < r_s$ for  $r_s\rightarrow 0$) with the
approximation $f(\nu,r_s\rightarrow 0)\approx \nu^{1/2}b_1$.
Using the more accurate function $f(\nu,r_s\rightarrow 0)\approx \nu^{1/2}
(1+1.99432r_s)+0.44713r_s$, where $c_{BG}^{3/2}/c_{WC}=0.44713$, does not change the expressions for
${\cal E}_{0 \, ,min}$ and $\nu_0$, but one needs to replace $b_1$ by $b_1=1+1.99432r_s$.
In the last expression for $f(\nu,r_s\rightarrow 0)$ the first term depends on $\nu$ and describes the effect
of statistics in ${\cal E}_{0 \, ,min}$. Without this term we had obtained (from the second term) the energy of
the 2DCBG. However, if we substitute the expression for $\nu_0$ with the new $b_1$ in
$f(\nu,r_s\rightarrow 0)$
and take the high density limit ($r_s\rightarrow 0$) then the first term of $f(\nu,r_s\rightarrow 0)$ is always
larger than the second one. This is the reason why we have a deviation from the energy of the 2DCBG. In Ref.~\ci{arm}
we have motivated the inclusion of the $\nu^{1/2}$ dependence in $f(\nu,r_s)$ by the linear $\nu$ dependence
of the ground state energy for the anyon gas without Coulomb interaction close to bosonic limit found in
\ci{sen},\ci{wen} and \ci{mori}. This linear dependence of the energy on $\nu$ is also obtained from Eq.~(38) of \ci{arm}
if we take the limit $r_s\rightarrow 0$ under the constraint $r_s < \nu$.

{\it In conclusion}, in the frame of our phenomenological approach based on the anyon concept, we have identified correlation
effects in the high density limit of the 2D Coulomb anyon gas. In contrast with results from MC calculations for fermions,
our data show a pronounced minimum in the correlation energy, which can be ascribed to bosonic correlations. Close to the
bosonic limit we find at finite $\nu$ a minimum of the ground state energy, which is lower than the known value for the 2DCBG
at $\nu = 0$ and is ascribed to statistical correlations. Neither of these effects is reported so far in
the literature.

\smallskip

{\it Acknowledgment}: The work was performed with support from the Volkswagen Foundation. B.~A. acknowledges the hospitality
at the University of Regensburg.

\end{document}